\documentclass[twocolumn]{svjour2}
\usepackage{graphicx}
\usepackage{amssymb}
\usepackage{mathptmx}      

\journalname{Journal of Statistical Physics, v. 133, p. 805 (2008)}

\begin{document}

\title{Field theory conjecture for loop-erased random walks}


\author{Andrei A. Fedorenko \and Pierre Le Doussal \and Kay J\"org Wiese }


\institute{A.A. Fedorenko \at
              CNRS-Laboratoire de Physique Th{\'e}orique de l'Ecole Normale Sup{\'e}rieure,
              24 rue Lhomond, 75231 Paris Cedex,  France \\
          \emph{Present address:} CNRS UMR 5672 - Laboratoire de Physique
            de l'Ecole Normale Sup{\'e}rieure de Lyon, 46 All\'{e}e d'Italie,
               69007 Lyon, France\\
          \email{andrey.fedorenko@ens-lyon.fr}
           \and
           P. Le Doussal  \at
      CNRS-Laboratoire de Physique Th{\'e}orique de l'Ecole Normale Sup{\'e}rieure,
              24 rue Lhomond, 75231 Paris Cedex,  France \\
              \email{ledou@lpt.ens.fr}
           \and
           K.J. Wiese  \at
            CNRS-Laboratoire de Physique Th{\'e}orique de l'Ecole Normale Sup{\'e}rieure,
              24 rue Lhomond, 75231 Paris Cedex,  France \\
              \email{wiese@lpt.ens.fr}
}

\date{Received: 21 May 2008 / Accepted: 23 October 2008 / Published online: 6 November 2008}

\maketitle

\begin{abstract}
We give evidence that the functional renormalization group (FRG), developed to study disordered systems, may
provide a field theoretic description for the loop-erased random walk (LERW), allowing to compute its
fractal dimension in a systematic expansion in $\varepsilon=4-d$. Up to two loop, the FRG agrees with
rigorous bounds, correctly reproduces the leading logarithmic corrections at the upper critical dimension $d=4$,
and compares well with numerical studies. We obtain the universal subleading logarithmic correction in $d=4$,
which can be used as a further test of the conjecture.
\end{abstract}


The loop-erased random walk (LERW) was introduced by Lawler \cite{lawler80} as an alternative to the self-avoiding
walk (SAW), which is relevant in polymer physics. On the lattice, the LERW is defined as the trajectory of a random walk in which any
loop is erased as soon as it is formed. As the SAW, the LERW has no self-intersections,  but is more tractable
mathematically. It has been proven  that the LERW has a scaling limit in all dimensions
\cite{duplantier92,kozma07,majumdar92,lawler95,lawler91,lawler99}. The
number of steps (or time) $t$
it takes to reach the distance $L$ scales as $t_L \sim L^{z}$, where $z$ is the fractal dimension of
LERW. In $d=2$ this scaling limit is conformally invariant and described \cite{schramm00,lawler04} by the stochastic
Loewner evolution SLE$_2$. Though the LERW and SAW belong to different universality
classes, the LERW itself has received significant attention due to applications in combinatorics, self-organized
criticality (SOC), and, more recently, conformal field theory and SLE. The LERW can be viewed as a special case
of the Laplacian random walk \cite{lawler87} and can be mapped \cite{majumdar92} to the problem of uniform spanning trees
(UST): Chemical (i.e.\ shortest) paths on UST obey LERW statistics. It is proven that the upper
critical dimension is $d_{\mathrm{uc}}=4$, the same as for the SAW, since for $d>4$ the traces of two random walks do almost surely not
intersect. Hence for $d>4$ the fractal dimension of the LERW is that of a simple random walk, $z=2$. It was proven  a while ago \cite{lawler88} that for $d < 4$ it is bounded from above by the Flory estimate for the
SAW exponent:
\begin{eqnarray}
R^2 \sim t^{2/z} \quad , \quad z < \frac{d+2}3=2-\frac{\varepsilon}3,  \label{bound}
\end{eqnarray}
where $R=\langle R(t)^2 \rangle^{1/2}$ is the radius of gyration, and we have introduced $\varepsilon=4-d$. The
mapping to UST and equivalently to the $q$-state Potts model at $q\to 0$ was used in $d=2$
to predict \cite{majumdar92} $z_{\mathrm{LERW}}(d=2)=\frac54$, later proved in Ref. \cite{Kenyon00}.
It is a particular case, for $\kappa=2$, of the fractal dimension $d_{\mathrm{f}}=1 +
\frac{\kappa}{8}$ of the trace of SLE$_\kappa$ (a simple curve for $\kappa<4$). Another connection in $d=2$ is to the
$O(N)$ loop model with $N=-2$, both corresponding to a conformal field theory (CFT) with central charge $c=-2$.
The leading logarithmic corrections at the upper critical dimension $d=4$ where
obtained  by Lawler \cite{lawler95} who proved that:
\begin{equation}
 R^2 \sim t(\ln t)^{1/3}. \label{log1}
\end{equation}
In $d=3$ the value of $z_{\mathrm{LERW}}$ is known only from numerics. The most precise estimate was obtained by
Agrawal and Dhar \cite{agrawal01},
\begin{equation}
  z_{\mathrm{LERW}}(d=3)=1.6183\pm 0.0004\ , \label{zd2}
\end{equation}
improving on the previous numerical result \cite{guttman90} $z=1.623 \pm 0.011$.

Contrary to the SAW, which is described by the $O(N)$ model at $N=0$, there seems to be at present no field-theoretic approach to compute the LERW exponent in a dimensional expansion around $d=4$.
This is surprising, especially when compared with the recent progress in CFT descriptions in $d=2$.
In this short paper we propose a field theoretic description for the LERW, based on the Functional RG, a method developed to
study disordered systems. We build on a connection proposed a
while ago between the depinning transition of periodic elastic systems, also called charge density waves (CDW) in
random media and sandpile models. The correspondence, on which we detail below,  is indirect, through a chain of related models: From LERW to UST to sandpiles to CDW-depinning and finally to functional renormalization group (FRG) field theory.  Some of the connections are
not rigorous. At the end we show that the FRG passes
all the tests of presently known results for LERW. In particular, it reproduces the correct leading logarithmic
corrections in $d=4$ given by Eq.~(\ref{log1}), and makes a prediction for the subleading logarithmic correction
which we hope will be tested in the near future.

We now present briefly the intermediate models. The Bak-Tang-Wiesenfeld (BTW) sandpile model
was proposed as a prototype for driven dissipative systems exhibiting SOC \cite{Bak87}. It is
defined on the $d$-dimensional hyper-cubic lattice with $L^d$ sites. The configuration at time
$t$ is given by the integer number of grains $h(x,t)$ at site $x$. The site $x$ is
unstable if $h(x,t)>2d$ in which case it relaxes
according to the toppling rule
\begin{eqnarray}
  h(x,t+1)&=&h(x,t)-2d\ , \nonumber \\
  h(y,t+1)&=&h(y,t)+1\ ,  \label{sandpile}
\end{eqnarray}
where $y$ denotes all the $2d$ nearest neighbors of site $x$. The neighbor sites may then
become unstable and the toppling continue. The order of topplings is irrelevant, thus the
notion of  ``Abelian'' sandpile, which allows to use e.g.\ parallel dynamics. The process
continues  until no unstable site remains, i.e.\ the avalanche ends. This is achieved
through grains that leave the system at the boundary. To ensure a steady state one drives
the system by adding a grain to a randomly chosen site $x$ after each avalanche. Stable
configurations are either transient, or recurrent in which case they appear in the steady
state with equal probability \cite{dhar90}.

Using the burning algorithm a one-to-one correspondence is set up between
the recurrent configurations of sandpile models and spanning trees on the same lattice
plus a sink $x_0$ (into which fall all grains leaving the system through the boundary)
\cite{majumdar-dhar92}. These spanning trees
are connected sets of edges touching all lattice sites with no loops, and chosen with uniform
probability (UST). Due to the Abelian property any avalanche initiated at an
arbitrary site $x_i$ can be decomposed into successive toppling waves, such that during one wave
each site topples at most once \cite{priezzhev96}. Each wave is characterized
by a two-root spanning tree on the extended lattice with connected $x_i$ and $x_0$. The subtree
with the root at $x_i$ spans the sites toppled by the wave while the subtree with the root
at $x_0$ connects the sites not affected by the wave. This spanning tree gives a
unique (chemical) path between two toppled sites separated by the distance $L$. The
number of edges along this path is nothing but the number of update steps between these
two toppling events, i.e. the time $t$ between them.
Thus the time $t$ and length  $L$ in the sandpile dynamics are related by the fractal dimension
of the chemical path along a spanning tree \cite{ktitarev00}, which is nothing but a LERW.

Narayan and Middleton proposed that the charge density wave (CDW) near the depinning transition
can be viewed to some extent as a BTW sandpile model \cite{narayan94,Alava2002}.
The configuration of a $d$-dimensional elastic object, such as a CDW, pinned by random
impurities is parameterized by a displacement field $u_{xt}$, $x \in \mathbb{R}^d$.
Here we assume that the distortions of the object caused by disorder are continuous.
For CDWs $u_{xt}$ is the local phase divided by $2\pi$.
In the continuous limit, the dynamics is described by the over-damped equation of motion
\begin{equation} \label{eq-motion}
\eta\partial_t u_{x t} = \nabla^2 u_{x t} + F(x,u_{x t})+f,
\end{equation}
where $\eta$ is the friction, $f$ the driving force, e.g. the external electric field for CDW,
and $F$ the impurity pinning force. The latter is taken to be Gaussian with zero mean and
correlator
\begin{eqnarray} \label{disorder-1}
\overline{F(x,u)F(x',u')}&=&{\Delta}(u-u')\delta^d(x-x'),
\end{eqnarray}
where $\Delta (u)$ is an even periodic function (chosen here with period 1).
For CDWs $F(x,u)$ is often written as
\begin{equation} \label{disorder-2}
F(x,u)=\alpha(x)Y[u-\gamma(x)]
\end{equation}
with $|Y(u)|\le1$ being a periodic function with period 1. The strength $\alpha(x)$ of impurities
and their selected phase $\gamma(x) \in [0,1)$ are taken to be random and
uncorrelated in space. Both forms (\ref{disorder-1}) and (\ref{disorder-2}) of disorder
correlations are equivalent at least in the vicinity of the so-called depinning transition.
The system  driven by a homogenous force $f$ undergoes
a depinning transition at a critical force $f_c$:
below the transition $f<f_c$ the system is pinned while above it, it exhibits a never-ending motion.
This dynamic transition is reminiscent of an ordinary second-order phase transition with the
velocity playing the role of the order parameter: $v\sim (f-f_c)^{\beta}$
and enjoys a similar universality, i.e.\ independence of the scaling behavior
on the microscopic details. The correlation length $\xi$ defined through the velocity-velocity
correlation function diverges at the transition as $\xi\sim (f-f_c)^{-\nu}$;  the length
and time scales of fluctuations are related by the dynamic critical exponent $z$:
$t\sim x^z$. Discretizing  space and time, one can rewrite
Eq.~(\ref{eq-motion}) near the depinning transition in the form of the automaton model
(\ref{sandpile}) with $h(x,t)$ playing  the role of a coarse-grained curvature of the elastic
object. For a pinning potential with narrow and steep wells the configuration of the CDW
can be encoded by integer variables $m(x,t)=u(x,t)-\gamma(x)$.
Upon time and space discretization $m(x,t)$ evolves according to the following
dynamic rule \cite{narayan94}
\begin{equation} \label{toppling2}
  m(x,t+1)=m(x,t)+\Theta\{f+\nabla^2[\gamma(x)+m(x,t)]+\alpha(x) \},
\end{equation}
where $\Theta(x)$ is the Heaviside step function and $\nabla^2$ is
the discrete Laplace operator. In terms  of a coarse-grained curvature
defined as
\begin{equation}\label{toppling3}
  h(x)= \sum\limits_{y}[m(y)-m(x)] + \textrm{int} [ f+\nabla^2\gamma(x)+\alpha(x) ],
\end{equation}
the dynamics rule (\ref{toppling2}) can be expressed in the form of the toppling rule
(\ref{sandpile}) with an inessential difference that the stability threshold now
is $h(x)=0$ not $h(x)=2d$. Increasing $f$ toward the critical force is equivalent to
increasing $h$ by 1 at a random site. Thus both models, the original BTW
model and the discretized version of Eq.~(\ref{eq-motion}), are driven by adding grains,
$h(x,t+1)=h(x,t)+1$.
In the CDW model grains are added with a cycle restriction: a second grain can be added
to a particular site only if all other sites have received a grain. The ordering of sites
in the cycle is provided by the value of the quenched random term $\nabla^2\gamma(x)+\alpha(x)$
in Eq.~(\ref{toppling3}).  While this difference between
the pinned at threshold CDW and BTW may seem inconsequential for large systems, it
illustrates that the mapping
is presently not rigorous; moreover an ad-hoc discretization was used. Nevertheless
it is supported by numerics \cite{narayan94}. It strongly suggests that the dynamic
exponent $z$ describing the depinning transition of a $d$-dimensional CDW coincides with
the fractal dimension of LERW in $d$ dimensions.

\begin{table}[tbp]
\caption{Pad\'{e} approximants  for $z$  in $d=3$.}
\label{table1}%
\begin{tabular}{l|llll}
\hline\noalign{\smallskip}
$[n/m]$   &    $m=0$   &  $m=1$    &   $m=2$      &  $m=3$     \\
\noalign{\smallskip}\hline\noalign{\smallskip}
 $n=0$    &    2       &  1.71429  &   1.6        &  1.61074   \\
 $n=1$    &  1.66667   &  1.5      &   1.61194    &            \\
 $n=2$    &  1.55556   &  1.63158  &              &            \\
 $n=3$    &  1.60069   &           &              &            \\
\noalign{\smallskip}\hline
\end{tabular}
\end{table}

The field theory which describes the depinning transition  is based on the
Functional RG \cite{depinning}. Applying the Martin-Siggia-Rose formalism to
Eq.~ (\ref{eq-motion}) and averaging over disorder with the help of Eq.~(\ref{disorder-1})
one arrives at the action
\begin{eqnarray}
S &=&\int_{x t} i\hat{u}_{x t}(\eta\partial_t - \nabla^2 )u_{x t}-
\int_{x t} i\hat{u}_{x t} f_{x t} \nonumber \\
& & -\frac12\int_{x t t'}\, i\hat{u}_{x t}i\hat{u}_{x t'}
\Delta({u}_{x t}-{u}_{x t'}),  \label{action}
\end{eqnarray}
where the response field $\hat{u}_{x t}$ has been introduced.
We argue that the field theory defined by (\ref{action}) also describes the LERW.
The dynamic exponent of model (\ref{action}) at criticality
is nothing but the fractal dimension of LERW. We also expect that more properties
of LERWs  are encoded in the fixed point $\Delta^*(u)$ since it controls the large-scale behavior of the model. At this stage, however, we only extract the relation
between time $t$ and space $L$, more work is needed to elucidate relations to other
observables.
Power counting shows that the upper critical
dimension of the model is $d_{\mathrm{uc}}=4$.
The peculiarity of the problem is
that all derivatives  of $\Delta(u)$ at the origin $u=0$ become relevant operators
below $d_{\mathrm{uc}}=4$. Thus, instead of a finite number of coupling constants
we have to deal with a whole coupling function $\Delta(u)$. To extract the scaling behavior
one has to study the flow of the renormalized function $\Delta(u)$  under
changing the infrared cutoff towards infinity. Recent progress has shown that full
consistency requires a 2-loop study. The corresponding flow equations to 2-loop order
read \cite{ledoussal02}
\begin{eqnarray}
  && \!\!\!\!\!\!\!\!
   \partial_l \Delta(u)= \varepsilon \Delta(u) -\frac12 [(\Delta(u)-\Delta(0))^2]^{\prime\prime}
        \nonumber \\
   &&  + \frac12 \left[\left(\Delta(u)-\Delta(0)\right)\Delta'(u)^2\right]^{\prime\prime}
 + \frac{1}2 [\Delta'(0^+)]^2\Delta''(u),\label{delta-flow} \ \\
  && \!\!\!\!\!\!\!\!
   \partial_l \ln \eta =
-\Delta''(0)+\Delta''(0)^2
+\Delta'''(0^+)\Delta'(0^+)\left[\frac32-\ln2 \right],\nonumber\\
 \label{eta-flow}
\end{eqnarray}
where $l =\ln L$, and $L$ the infrared cutoff, e.g.\ the size of the system. Below $d=4$ the flow equation (\ref{delta-flow}) has a fixed point (FP) solution $\Delta^{*}(u)$ with a cusp
at the origin: $\Delta^{*\prime}(0^+)\ne 0$. Taking into account that $t \sim \eta_l L^2$ and that in the vicinity
of the FP $\eta_l$ scales with $L$ according to Eq.~(\ref{eta-flow}), one finds that $t \sim L^{z}$, and to
2-loop order the exponent $z$ is given by \cite{ledoussal02}:
\begin{eqnarray}\label{z}
  z=2-\frac{\varepsilon}3-\frac{\varepsilon^2}9+O(\varepsilon^3).
\end{eqnarray}
Comparison with the exact bound (\ref{bound}) is encouraging since the 2-loop correction has the correct
sign. To estimate values of $z$ in $d=2,3$ we compute different Pad\'{e} approximants $[n/m]$ to (\ref{z}). In
$d=2$ we obtain $z(d=2)=1.23 \pm 0.22$, consistent with the presumed exact value $\frac54$ but
of poor accuracy \footnote{We used the average and root mean squared of the set $z_{[1/0]}=4/3$, $z_{[0/1]}=3/2$, $z_{[2/0]}=8/9$, $z_{[0/2]}=6/5$. We excluded approximant $[1/1]$ which
accidentally vanishes.} due to the large expansion parameter
$\varepsilon=2$. Table \ref{table1} contains the approximants $[n/m]$
in $d=3$. To improve the accuracy we construct the higher order Pad\'{e} approximants with $n+m=3$
imposing that $z(d=2)=\frac54$ for $\varepsilon=2$. The average and root-mean-square of all  approximants with $n+m=3$
yields
\begin{eqnarray} \label{zest}
z= 1.614 \pm 0.011\ .
\end{eqnarray}
The same procedure based on one loop only produces $z= 1.638 \pm 0.012$. The value (\ref{zest}) is our
best 2-loop FRG prediction for $d=3$ and is in fairly good agreement with the numerical result (\ref{zd2}).

The field theory is most predictive at the upper critical dimension,
where it yields exact results. For FRG some were obtained
previously (see e.g.\ Refs.\ \cite{chitra,fedorenko03,higherorder}).
Here we compute the logarithmic corrections to the dynamics
which yield a prediction for the LERW in $d=4$.
For $\varepsilon=0$ one shows from (\ref{delta-flow}) that the (periodic) disorder correlator approaches
the FP solution $\Delta^*(u)=0$ as follows (for $0 \leq u \leq 1$):
\begin{equation} \label{delta-d4}
  \Delta_l(u)=\left[\frac1{6l}+\frac{\ln l}{9l^2}\right]
  \left[\frac16-u(1-u)\right]+O\!\left(\frac1{l^2}\right).
\end{equation}
Substituting Eq.~(\ref{delta-d4}) into Eq.~(\ref{eta-flow}) we obtain
\begin{equation}
  \ln \frac{\eta_l}{\eta_0} = - \frac1{3}\ln l +\frac{2\ln l}{9l}+O\!\left(\frac1{l}\right).
\end{equation}
Renormalizing the relation  $t\sim \eta_l L^2$ up to scale $L$ we arrive at
\begin{equation}
  t\sim L^2 (\ln L)^{-1/3}
  \left[1+\frac{2 \ln\ln L}{9\ln L}  + O\!\left(\frac1{\ln L}\right)\right],
\end{equation}
which can be rewritten as
\begin{equation} \label{loglog}
  L^2 \sim t (\ln t)^{1/3}
  \left[1-\frac{\ln\ln t}{3\ln t}  + O\!\left(\frac1{\ln t}\right)\right].
\end{equation}
The scale $L^2$ can be taken as $R^2$, the radius of gyration. We note that
the leading order of Eq.~(\ref{loglog}) coincides with the result (\ref{log1}) of
Lawler. Here we obtain the universal subleading correction.

The prediction (\ref{loglog}) could be tested numerically, very much as for the
corresponding prediction \cite{duplantier} for the SAW in $d=4$, checked in  \cite{grassberger}; there subleading corrections are necessary to properly fit the numerical data at any feasible chain length. To this purpose, it is useful to note that, as for the SAW, there is only a single non-universal constant in the correction term in the parenthesis in (\ref{loglog}), $c/\ln t$. It can  e.g.\ be put to zero by a proper choice of $t_0$, setting $t \to t/t_0$ (in
which case further corrections are $1/\ln^2 t$ and determined by 3- and higher-loop terms, not
considered here). For comparison, one similarly finds, from the 2-loop $\beta$-function \cite{ZinnBook} of the
$O(N)$ model in $d=4$ that
$L^2 \sim t (\ln t)^{a_N} (1-b_N \frac{\ln\ln t}{\ln t})$ with $a_N=(N+2)/(N+8)$ and $b_N=(N+2)(68+8N-N^2)/(N+8)^3$
from which the Duplantier values for the SAW \cite{duplantier}, $a_0=1/4$ and $b_0=17/64$ are retrieved at $N=0$.
Note that no value of $N$ can account for (\ref{loglog}), thus a representation of LERW via the $O(N)$ field theory, if feasible at all, would at least require a more complicated operator correspondance.

Note finally that the  exponent $\tau$ for the avalanche-size distribution \footnote{The probability to have  an avalanche of size $s$ is proportional to $s^{-\tau}$, with a cutoff for large and small avalanches. }
was recently computed to one loop within FRG \cite{statics,dynamics}. In the sandpile literature no controlled
calculation exists for the corresponding exponent, usually called
$\tau_s$, but the formula $\tau_s(d)=2-2/d$ leading to $\tau_s=4/3$ in $d=3$
has been conjectured \cite{agrawal01} from scaling arguments. The FRG was found to
agree to $O(\epsilon)$ with this formula, assuming $\tau=\tau_s$. Evaluation of 2-loop corrections is
in progress as a further test of the conjecture and of the relations between
sandpile models and depinning.

To conclude, it may appear surprising that LERW be related to a field theory based on functional RG, where
one would expect a ``simpler'', more conventional, field theory based on a single relevant
coupling constant, as for the SAW.
One clue may be that for periodic systems (CDW) the FRG posesses a
stable submanifold with only two coupling constants, $\Delta(u)=a+ b u(1-u)$ (for $0 \leq u \leq 1$),
which contains the leading critical behavior. From (\ref{delta-flow})
one obtains \cite{ledoussal02,dynamics}
the spectrum of convergence to the fixed point \footnote{Apart from the trivial uniform mode
of eigenvalue $\epsilon$.} as $1/(\ln L)^{1+\alpha_n}$ in
$d=4$ and $L^{-\omega_n}$ with $\omega_n= \alpha_n \epsilon + \beta_n \epsilon^2 + O(\epsilon^2)$ for $d<4$;
$\alpha_n = \frac13  (3 + n) (1 + 2 n)$, $ \beta_n = -\frac{1}{9} (n+2) (2 n+1) (2 n+3)$.  The values $n=1,2,\ldots$ correspond to the convergence to the submanifold,
and $n=0$ to the convergence inside it. Convergence to the submanifold is fast, with
leading eigenvalue $1+\alpha_1=5$. Fast convergence was also found  in numerics \cite{alan} where
the FRG function $\Delta(u)$ in (\ref{delta-flow}) was directly {\it measured} in $d=1,2,3$.
The conjecture raises many other interesting issues to be explored, such as
the possibility to predict other LERW observables and corrections to scaling, the role of other universality classes for CDW
and interface depinning, the connections to fermionic field theory (known for UST, see e.g.\ \cite{Read04},
and conjectured for the FRG \cite{higherorder}). Work is in progress in these directions.

\begin{acknowledgements}
We acknowledge support from the Agence Nationale de la Recherche
under program 05-BLAN-0099-01 and from the European Commission under contract
No.~MIF1-CT-2005-021897 (AAF).
\end{acknowledgements}

\end{document}